\documentstyle[amssymb,epsfig,aps,prl,psfig,floats]{revtex}

\begin{document}

\twocolumn[
\hsize\textwidth\columnwidth\hsize\csname
@twocolumnfalse\endcsname

\title{Beliaev damping of quasi-particles in a Bose-Einstein condensate}
\author{N. Katz, J. Steinhauer, R. Ozeri, and N. Davidson}
\address{Department of Physics of Complex Systems,\\
Weizmann Institute of Science, Rehovot 76100, Israel}
\maketitle

\begin{abstract}
We report a measurement of the suppression of collisions of
quasi-particles with ground state atoms within a Bose-Einstein
condensate at low momentum. These collisions correspond to Beliaev
damping of the excitations, in the previously unexplored regime of
the continuous quasi-particle energy spectrum. We use a
hydrodynamic simulation of the expansion dynamics, with the
Beliaev damping cross-section, in order to confirm the assumptions
of our analysis.
\end{abstract}
]

In a Bose-Einstein condensate (BEC) collisions with
distinguishable excitations (impurities) and collisions with
indistinguishable excitations (quasi-particles) differ profoundly.
These differences follow from the quantum exchange symmetry of the
indistinguishable excitations, and from a different excitation
spectrum.

Impurity collisions within a BEC have been measured previously.
Using Raman spectroscopy the microscopic onset of superfluidity
was measured and found to be in general agreement with prediction
\cite{ketterle-impure}. The collisional dynamics and interaction
between two distinguishable slowly moving condensates was measured
\cite{modugno} and found to agree with
simulation\cite{Band-collisions}. Macroscopic superfluid behavior
has also been demonstrated, involving the interaction of the
condensate with large-scale optical dipole-potential structures
\cite{Ketterle-raster}, \cite{Italy-sloshing}.

The case of identical particle collisions has been extensively
studied using many-body theory, starting with \cite{beliaev}.
Recently, these results have been applied to BEC explicitly
\cite{landau-stringari}, \cite{damping-giorgini}, \cite{decay
rates}, \cite{griffin}. In this letter we present a measurement of
collisions between quasi-particles and the BEC, at velocities near
and above the superfluid critical velocity $v_{c}$.

According to the Fermi golden rule, the rate of scattering within
a homogenous BEC is given by \cite{ketterle-review}:
\begin{equation}
n \sigma _{k} v_{k} =  2^{7/2}n a^{2} v_{c} \int dqd\Omega
q^{2}\left| A_{q;k}\right| ^{2} \delta (E_{i}-E_{f})
\label{sup-general}
\end{equation}
where $a$ is the s-wave scattering length. The wavenumber $k$ of
the excitation is in units of $\xi ^{-1}=\sqrt{8\pi na}$, the
inverse healing length of the condensate. The free particle
velocity of the excitations is $v_{k}=\hbar k \xi ^{-1} /m$, where
$m$ the mass of the BEC atoms. $n$ is the density of the
condensate. The superfluid critical velocity is
$v_{c}=\sqrt{\mu/m}$, where $\mu =gn$ is the chemical potential of
the BEC, and $g$ is $4\pi \hbar^{2}a/m$. The integral is over all
possible momentum transfers $q$, and all angles $\Omega$. The
$\delta$-function requires energy conservation between the initial
energy $E_{i}$ and the final energy $E_{f}$ after collision, where
we express energy in units of $\mu$. The factor $A_{q;k}$ is the
$q$-dependent, momentum conserving, scattering matrix element of
an excitation of wavenumber $k$, which may include suppression or
enhancement of the collision process due to many-body effects.

For impurity scattering within a condensate
\cite{ketterle-review}, $\left| A_{q;k}\right|^{2}$ is given by
the structure factor $S_{q}$ \cite{Pitaevskii}. The initial energy
$E_{i}=E_{k}^{0}$, where $E_{k}^{0}=k^{2}$ is the impurity
dispersion relation, for impurities with mass $m$. The final
energy for this process is $E_{f}=E_{\mathbf{k-q}}^{0}+E_{q}^{B}$,
where $E_{q}^{B}=\sqrt{q^{4}+2q^{2}}$ is the recently measured
\cite{Steinhauer} Bogoliubov dispersion relation \cite{Bugo47}.

Below $v_{c}$ collisions between the impurity and condensate are
completely suppressed (see Fig. 1) by the $\delta$-function
requiring conservation of energy and momentum.

\begin{figure}[h]
\begin{center}
\mbox{\psfig{figure=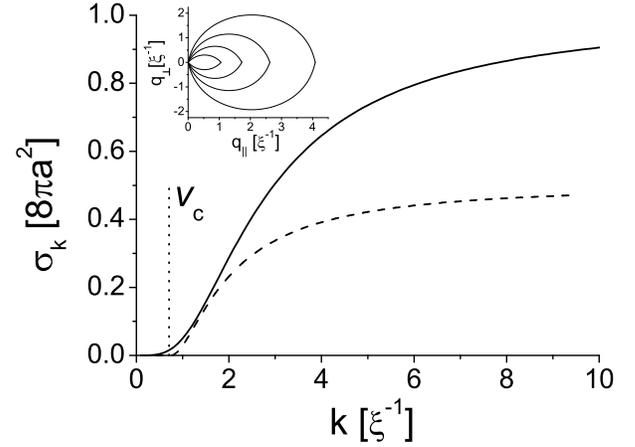,width=8cm}}
\end{center}
\vspace{0.4cm} \caption{Cross-section for collisions in a
homogenous condensate. The momentum is in units of the inverse
healing length, $\xi ^{-1}=\sqrt{8\pi na}$. The cross-section is
in units of the free particle scattering cross-section for
identical particles $8\pi a^{2}$. The solid line is the
theoretical curve, from Eq. (3) for quasi-particles travelling
through the condensate. The dashed line is the cross-section for
impurity scattering within the condensate [1], which vanishes
below the superfluid critical velocity $v_{c}$. The inset shows
the allowed momentum manifold for quasi-particle collisions due to
conservation of energy and momentum. The $q_{\|}$ and $q_{\bot}$
axes correspond to the parallel and orthogonal components of the
scattered momentum respectively, where
$tan(\theta)=q_{\bot}/q_{\|}$. The manifolds represent the
experimental $k$'s 4.09 (outermost), 2.63, 1.73 and 1.06
(innermost).}
\end{figure}

In the case of quasi-particle collisions only the Bogoliubov
dispersion relation is relevant, and the $\delta $-function in Eq.
(\ref{sup-general}) is simply $\delta
(E_{k}^{B}-E_{q}^{B}-E_{\mathbf{k}-\mathbf{q}}^{B})$. We solve
this condition, and find the angle $\theta $ between
the initial direction $\mathbf{k}$ and the scattered direction $%
\mathbf{q}$, to be (see inset of Fig. 1):

\begin{equation}
\cos (\theta )=(2kq)^{-1}\left[ k^{2}+q^{2}+1-\sqrt{%
1+(E_{k}^{B}-E_{q}^{B})^{2}}\right]  \label{sup-angle}
\end{equation}

This result differs in a qualitative way from impurity scattering,
since Eq. (\ref{sup-angle}) has solutions for any finite $k$.
There is no longer any well-defined critical velocity at which
collisions are completely suppressed. However, not all angles are
allowed. At a given $k$ we find that the maximal allowed angle is
$cos(\theta_{max})=\sqrt{(k^2+2)/2}/(k^2+1)$. At the limit of
small $k$, this angle approaches zero, and collisions are allowed
only for $\mathbf{q}$ parallel to $\mathbf{k}$.

The appropriate suppression term $\left| A_{q;k}\right| ^{2}$ for
quasi-particles has been calculated \cite{landau-stringari},
\cite{damping-giorgini}. In this work we expect mainly Beliaev
processes which involve creation of lower energy excitations. The
Landau damping rate is expected to be an order of magnitude slower
than the observed Beliaev collision process
\cite{damping-giorgini}.

We start with the atomic interaction Hamiltonian $H^{^{\prime }}=
\frac{g}{2V} \sum_{j,l,m,n}a_{j}^{\dagger
}a_{l}^{\dagger }a_{m}a_{n}\delta _{j+l-m-n}$, where $V$ is the volume of the BEC, $a_{i}^{\dagger}$ and $a_{i}$ are
the atomic creation and annihilation operators at wavenumber $i$. We approximate $%
a_{0}^{\dagger }\thickapprox a_{0}\thickapprox \sqrt{N_{0}}$, with
$N_{0}$ the number of atoms in the condensate. We take the
Bogoliubov transform
$a_{p}^{\dagger }=(u_{p}b_{p}^{\dagger }-v_{p}b_{-p})$, with $u_{p}$ and $%
v_{p}$ the appropriate quasi-particle amplitudes, which were
recently
measured \cite{ketterle-bogo}. We are interested in terms of the form $%
b_{k}b_{\mathbf{k}-\mathbf{q}}^{\dagger }b_{q}^{\dagger }$, that
remove a quasi-particle of wave number $k$, and create two in its
stead. Calculating the matrix element prefactor of this term in
the atomic interaction
Hamiltonian, we arrive at $A_{q;k}=\frac{1}{2}%
(S_{q}+3S_{q}S_{k}S_{\mathbf{k}-\mathbf{q}}+S_{%
\mathbf{k}-\mathbf{q}}-S_{k})/\sqrt{S_{k}S_{q}S_{%
\mathbf{k}-\mathbf{q}}}$. This result can be viewed as the
explicit zero temperature limit of more general calculations
\cite{griffin}.

Applying Eq. (\ref{sup-angle}) and $\left| A_{q;k}\right| ^{2}$ to
Eq. (\ref{sup-general}),
and using the Feynman relation \cite{feynmann} $%
S_{q}=E_{q}^{0}/E_{q}^{B}$,\ we arrive at the rate of
excitation-condensate collisions:

\begin{equation}
n\sigma _{k}^{B}v_{k}=8\pi na^{2}v_{k}\times \frac{1%
}{2k^{2}}\int\limits_{0}^{k}dqq\left| A_{q;k}\right| ^{2}\frac{%
E_{k}^{B}-E_{q}^{B}}{\sqrt{1+(E_{k}^{B}-E_{q}^{B})^{2}}}
\label{sup-rate}
\end{equation}

The effective cross-section $\sigma_{k}^{B}$ for the
quasi-particles, is shown in Fig. 1 (solid line). For large $k$,
$\sigma _{k}^{B}$ approaches $8\pi a^{2}$, compared to $4\pi
a^{2}$ for impurities (dashed line). This enhancement by a factor
of 2 is due to the boson quantum mechanical exchange term. In Eq.
(\ref{sup-rate}), for small $k$, we verify that the scattering
rate indeed scales as $k^5$, which is the classic result
\cite{beliaev}, \cite{damping-giorgini}. In particular, it remains
finite even for $v_{k}<v_{c}$, in contrast with impurity
scattering.

In \cite{ketterle-impure}, the identical particle collision cross
section for large $k$ was measured to be $2.1(\pm 0.3)\times 4\pi
a^{2}$. Scattering rates in four-wave mixing experiments in BEC
\cite{four-wave} were also shown to agree with the high-$k$ limit
of Eq. (\ref{sup-rate}) \cite{band-collisions prl}.

In the opposite regime of extremely low wavenumber, where the
energy levels are discrete, Beliaev damping was observed for the
scissors mode of a BEC \cite{beliaev-observe}. The discrete energy
levels were tuned so that Beliaev damping of the initial mode to
exactly one mode of half the energy was achieved. Eq.
(\ref{sup-rate}) did not apply, since there was no need to
integrate over various scattering modes.

Our experimental apparatus is described in \cite{Steinhauer}.
Briefly, a nearly pure ($>95\%$) BEC of $10^{5}$ $^{87}Rb$ atoms
in the $|F,m_{f}\rangle =|2,2\rangle $ ground state, is formed in
a QUIC\ type magnetic trap \cite{QUIC}. The trap is cylindrically
symmetric, with radial ($\hat{r}$) and axial ($\hat{z}$) trapping
frequencies of $2\pi \times 220$ Hz and $2\pi \times 25$ Hz,
respectively. Thus $\xi =0.24\mu m$ via averaging in the local
density approximation (LDA) \cite{Steinhauer}.

\begin{figure}[h]
\begin{center}
\mbox{\psfig{figure=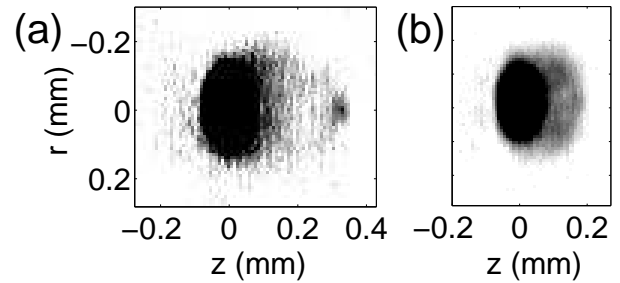,width=8cm}}
\end{center}
\vspace{0.4cm} \caption{Absorption TOF images of excited
Bose-Einstein condensates. \ (a) Absorption image for $k=2.63$,
with the large cloud at the origin corresponding to the
unperturbed BEC. A clear halo of scattered atoms is visible
between the BEC and the cloud of unscattered outcoupled
excitations. (b) Absorption image for $k =1.06$. For this value of
$k$ the distinction between scattered and unscattered excitations
is not clear, since both types of excitations occupy the same
region in space.}
\end{figure}

We excite quasi-particles at a well-defined wavenumber using
two-photon Bragg transitions \cite{Phillips-Bragg}. The two Bragg
beams are detuned 6.5 GHz from the $5S_{1/2},F=2\longrightarrow
5P_{3/2},F^{\prime }=3$ transition. The frequency difference
$\Delta \omega $ between the two lasers is controlled via two
acousto-optical modulators. Bragg pulses of 1 msec duration are
applied to the condensate. The angle between the beams is varied
to produce excitations of various $k$ along the z-axis, thus
preserving the cylindrical symmetry of the unperturbed BEC. The
beam intensities are chosen to excite no more than 20\% of the
total number of atoms in the condensate.

After the Bragg pulse, the magnetic trap is rapidly turned off,
and after a short acceleration period the interaction energy
between the atoms is converted into ballistic kinetic energy
\cite{Tomography}. After 38 msec of time-of-flight (TOF) expansion
the atomic cloud is imaged by an on-resonance absorption beam,
perpendicular to the z-axis. Fig. 2a shows the resulting
absorption image for $k=2.63$, with the large cloud at the origin
corresponding to the BEC. A halo of scattered atoms is visible
between the BEC and the cloud of unscattered outcoupled
excitations. No excitations with energy greater than that of the
unscattered excitations are observed, confirming our low estimate
of the Landau damping rate. Fig. 2b shows the absorption image for
$k =1.06$. For this $k$ value the distinction between scattered
and unscattered excitations is not clear in the image, since both
types of excitations occupy the same region in space.

At a given $k$ the number of excitations is varied by scanning
$\Delta \omega $ around the resonance frequency $\omega_{k}^{B}$.
The number of excitations $N_{mom}$ is measured by determining the
total momentum (in units of the recoil momentum $\hbar k
\xi^{-1}$) contained in the outcoupled region outside the
unperturbed BEC. This region includes all the scattered and
unscattered excitations, in the direction of $\mathbf{k}$. Thermal
effects are removed by subtracting the result of an identical
analysis over the other side. The results are shown in Fig. 3a.

\begin{figure}[h]
\begin{center}
\mbox{\psfig{figure=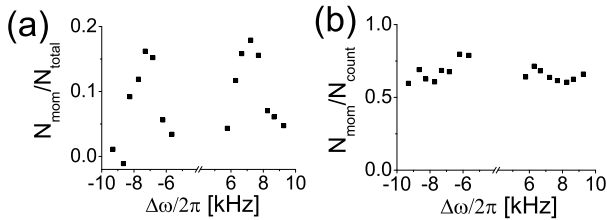,width=8cm}}
\end{center}
\vspace{0.4cm} \caption{Quantifying the amount of collisions
measured for $k=2.63$. \ (a) Measured momentum of the outcoupled
atoms vs. $\Delta \omega$. The momentum is in units of the recoil
momentum, and is normalized by the total number of atoms. \ (b) \
The measured ratio between $N_{mom}$ and $N_{count}$ (both defined
in the text). Every collision outcouples more atoms, increasing
$N_{count}$ but leaving $N_{mom}$ almost unchanged.}
\end{figure}

In order to quantify the amount of collisions, despite the lack of
separation between scattered and unscattered excitations, we take
the ratio between $N_{mom}$ and the counted number of atoms
$N_{count}$, in the same region. The resulting
$N_{mom}/N_{count}$, as a function of $\Delta \omega$, are shown
in Fig. 3b. The ratio is seen to be independent of the number of
excitations, and appears to be an intrinsic property of a single
excitation. Each collision between an excitation and the
condensate creates an additional excitation that is counted in the
outcoupled region, increasing $N_{count}$, while
the momentum ($N_{mom}$) in this interaction is conserved. Thus the ratio, $%
N_{mom}/N_{count}$, is a good quantifier of the amount of the
collisions, even at low $k$ \cite{comment-tomography}. Bosonic
amplification of the collision rate \cite{ketterle-review}, which
would appear as minima in Fig 3b, is not observed.

At a given $k$ we define the overall probability for an excitation
to undergo the first collision $p_{k}$. If we ignore secondary
collisions the result is $N_{mom}/N_{count}=1/(1+p_{k})$, since
each collision outcouples, after TOF, an additional particle
\cite{comment3}. Using this relation we infer $p_{k}^{exp}$ for
the various measured $k$'s.

We expect the scattering probability $p_{k}$ to be equal to
$\tilde{n} \sigma_{k}^{B} v_{k} t_{eff}$, where $t_{eff}$ is the
effective interaction time of the excitation with condensate and
$\tilde{n}$ is the average density.

We assume $t_{eff}$ to be $k$-independent, divide $p_{k}$ by
$v_{k}$, and arrive at a value that is proportional to the
scattering cross section (since the $\tilde{n}$ is constant for
all $k$). This assumption will be tested below, but must be valid
for sufficiently low $k$, for which the TOF expansion lowers the
density rapidly, turning off collisions before the excitations
move significantly.

The ratio, $p^{exp}_{k}/(\tilde{n} v_{k} t_{eff})$ is shown in
Fig. 4 ($\blacksquare$) and seen to agree with the theoretical
suppression (solid line) calculated as a LDA average
\cite{Pitaevskii} of Eq. (\ref{sup-rate}) \cite{comment2}. Since
$t_{eff}$ is not known, absolute calibration is not possible.
Therefore, the results are shown with arbitrary units.

In order to verify the validity of these simplifying assumptions,
$3$-D simulations are performed, using the hydrodynamic
Gross-Pitaevskii equations \cite{dalfovo-stringari}. The
simulations \cite{Tomography} include the TOF dynamics. The
expansion of the main BEC cloud is taken from expressions in
\cite{Castin+Dum}, computed for an elongated condensate by the
hydrodynamic equations. The excitations travel within the time
varying mean field potential created by the expanding unperturbed
condensate. The excitations collide with the expanding condensate
with the correct local scattering cross section, including the
angular dependence, taken from Eq. (\ref{sup-rate}). The resulting
distribution of simulated outcoupled atoms is analyzed by the same
method as the experimental data.

\begin{figure}[h]
\begin{center}
\mbox{\psfig{figure=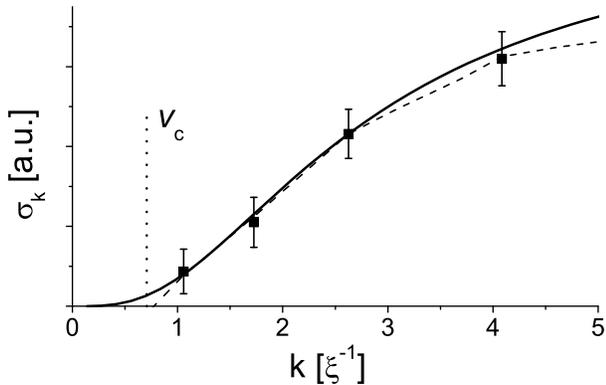,width=8cm}}
\end{center}
\vspace{0.4cm} \caption{Suppression of identical particle
collisions. Scattering cross-section is in arbitrary units. The
momentum is in units of the inverse healing length after LDA
averaging, $\xi=0.24\mu m$. The theoretical curve (solid line) is
a LDA average of Eq. (3) [25]. The assumptions of our analysis are
tested using hydrodynamic simulations, and found to agree with
Beliaev damping theory in the experimental regime (dashed line).}
\end{figure}

We plot the simulated $p_{k}/v_{k}$ (dashed line, in the same
units as experiment). The simulation and Beliaev damping theory
agree in the regime of experimentation, indicating a window of
validity for the assumptions used in analyzing the experimental
data. At large $k$ (above $k=4.09$), $t_{eff}$ is made shorter by
the rapid transit of the condensate by the excitations. At low $k$
(less than $k=1.06$) many of the collisional products remain
inside the condensate volume, and are not counted, preventing
analysis in this regime.

The arbitrary unit suppression used in Fig. 4 (both for simulation
and experiment) was set at $k=2.63$. The absolute cross-sections
obtained by comparing the experimental data to the hydrodynamical
simulations were higher than the theoretical values by a
k-independent overall factor of $2.36 \pm 0.08$, which is not
understood. This factor may be caused by various inaccuracies in
the TOF parameters of the simulation. However, the trend in the
experimental analysis is robust, and does not depend on absolute
calibration.

We also set the collision rate artificially to zero in the
simulation, and find $N_{mom}/N_{count}$ to be unity within $2\%$,
for all $k$. This implies that there are no significant other
mean-field repulsion effects along the $z$-axis \cite{Tomography},
confirming our assumption of momentum conservation.

In conclusion, we report a measurement of the suppression of the
collision cross-section for identical particles within a
Bose-Einstein condensate. We find the suppressions in our
experiment in agreement with a calculation of Beliaev damping
rates, within an overall factor. We use a hydrodynamic simulation
of the expansion dynamics, in order to verify our analysis of the
experiment. This represents the first measurement of this effect
in the quasi-particle continuous spectrum regime.

In the low temperature limit discussed in this Letter, the
non-linear coupling mechanism of the interaction Hamiltonian is
predicted to generate squeezing and entanglement of quasi-particle
excitations \cite{squeezing}. Further experimental study should
also lead to an understanding of these processes in the context of
the decoherence of excitations and coherent states due to
collisional losses within a BEC \cite{griffin-thermal relaxation},
\cite{wieman damping}, \cite{italy-atomoptic}.

This work was supported in part by the Israel Science Foundation
and by Minerva foundation.


\begin{references}
\bibitem{comment-tomography} For $k =1.06$, there are atoms in front and
behind the BEC that should be counted in our integration. We use
computerized tomography \cite{Tomography}, to find the correct
atomic distribution. The systematic error caused by naively
integrating the absorption image, can
be as large as $10\%$, which is unacceptable for our purposes. At higher $%
k$, these systematic integration effects are verified to be
negligible.
\bibitem{comment3} hydrodynamical simulations (described below) indicate that there are few higher order multiple
collisions for the experimental parameters. We also find that
nearly all the collision products are indeed located in the
counting region after TOF, confirming our momentum and atom number
counting procedures.
\bibitem{comment2} The LDA average of the suppression of collisions for an non-homogenous
BEC is nearly indistinguishable from the suppression of a
homogenous condensate with the LDA average healing length as
defined in \cite{Steinhauer}.
\end{references}
\end{document}